\documentclass[epj]{svjour}
\usepackage{graphics}
\begin{document}
\title{Nonmonotonic inelastic tunneling spectra due to surface spin excitations in ferromagnetic junctions}
\author{G. Tkachov
}                     
%
%
\institute{Institute for Theoretical Physics, Regensburg University, 93040 Regensburg, Germany\\
Institute for Theoretical Physics and Astrophysics, W\"urzburg University, Am Hubland, 97074 W\"urzburg, Germany}
\date{Received 5 April 2011 / Received in final form 3 July 2011 }
%
\abstract{
The paper addresses inelastic spin-flip tunneling accompanied by surface spin excitations (magnons) in ferromagnetic junctions. 
The inelastic tunneling current is proportional to the magnon density of states which is energy-independent for the surface waves 
and, for this reason, cannot account for the bias-voltage dependence of the observed inelastic tunneling spectra. 
This paper shows that the bias-voltage dependence of the tunneling spectra can arise from the tunneling matrix elements of the electron-magnon interaction. 
These matrix elements are derived from the Coulomb exchange interaction using the itinerant-electron model of magnon-assisted tunneling. 
The results for the inelastic tunneling spectra, based on the nonequilibrium Green's function calculations, are presented for both parallel and antiparallel 
magnetizations in the ferromagnetic leads.  
} 
\maketitle
\section{Introduction}
\label{sec_intro}

Spin polarized transport in tunnel ferromagnetic junctions 
has been a subject of intense research motivated by the desire to develop a form of electronics which utilizes the
dependence of the junction resistance on the carrier spin-polarization~\cite{Pri95}. 
Since ferromagnetic metals have more band electrons of one spin polarization
(known as majority carriers) present at the Fermi energy $E_F$
than of the inverse polarization (minority carriers), the resistance 
depends on the relative orientation of the magnetic moments in the ferromagnets 
which is controlled by an external magnetic field. 
With parallel magnetizations, the tunneling occurs between majority (and minority)
bands whereas in a junction with antiparallel magnetizations 
carriers tunnel from majority to minority bands (and vice versa). 
The resulting spin current mismatch produces a larger contact resistance in the antiparallel case, 
an effect known as the tunneling magnetoresistance (TMR)~\cite{Jul75,Miyazaki95,Moo95,Gall97,Sun98,Moo98,Moo00,Jap1,Jap2,Jap3,Kreuzer02,Kreuzer03,Zenger04}. 

Among various studies of the TMR effect a large body of work has aimed at developing  
theoretical approaches to spin-dependent tunneling in the single-particle approximation
~\cite{Slonczewski89,Bratkovsky97,MacLaren97,Mavr00,Mathon01,Butler01}, 
including many-body spin-dependent phenomena~\cite{Zhang97,MacDonald98,Guinea98,Brat98,Vedyayev01,GT02a,Hong02,GT02b,McCann02,McCann03,GT03}  
and effects of disorder~\cite{Tsymbal98,Tsymbal03,Mathon06,Xu06,GT08}.

This paper considers inelastic tunneling processes accompanied by a spin-wave excitation (magnon) in a biased ferromagnetic junction.  
Such processes become relevant at bias voltages, $V$, of order of {\it hundred} millivolts~\cite{Moo95,Gall97,Zhang97,Sun98,Moo98},  
which corresponds to typical spin-wave energies and the Curie temperatures, $T_C$, of commonly used ferromagnets such as Co or Ni$_{80}$Fe$_{20}$.  
The magnon-assisted tunneling involves an electron spin-flip and, for this reason, reduces the resistance for the antiparallel ferromagnet alignment,
which, in turn, results in a decrease in the TMR~\cite{Zhang97,Brat98}. 
Experimentally, the inelastic contribution to the tunneling current, $I(V)$, can be identified by taking the second derivative $d^2I/dV^2$ 
(see, e.g. Refs.~\cite{Moo98,Moo00}). It is related at low temperatures to the magnon density of states (MDOS) $\Omega$ at energy $|eV|$ 
(see, e.g. Refs.~\cite{Brat98,Guinea98,GT02a}): 
\begin{eqnarray}
d^2I/dV^2\propto {\rm sign}(V)\Omega(|eV|),
\label{simpleIETS}
\end{eqnarray} 
where $e$ is the electron charge. 
The reason for taking the second derivative of $I(V)$ is due to the fact that 
the inelastic current involves two integrations, over the energy of the tunneling electron and 
over the magnon energy, both limited by $eV$.   

From Eq. (\ref{simpleIETS}) one can draw the following conclusions. 
For bulk magnons with the usual quadratic dispersion $\omega_{\bf q} \propto {\bf q}^2$ 
(where ${\bf q}$ is the {\em three}-dimensional bulk magnon wave vector), 
the MDOS is proportional to the square root of the magnon energy, $\Omega\propto \omega^{1/2}$, 
and the magnon-assisted contribution vanishes in the zero-bias limit as $d^2I/dV^2\propto |V|^{1/2}$. 
In contrast, for surface magnons with similar dispersion 
$\omega_{{\bf q}_\|} \propto {\bf q}^2_\|$, 
propagating in the contact plane with a {\em two}-dimensional wave vector ${\bf q}_\|$, 
the MDOS is energy-independent, $\Omega =const$, and so is the second derivative $d^2I/dV^2\propto const$. 
The latter conclusion cannot, however, be true because for $V=0$ there is no extra energy to be transferred 
to the collective excitations and, therefore, the inelastic contribution must vanish. 
This apparent contradiction is believed to arise because Eq. (\ref{simpleIETS}) does not include 
the matrix elements of the electron-magnon interaction 
which should also depend on the bias voltage. Although this might be the valid explanation, 
it has not been supported yet by the direct calculation of the inelastic tunneling spectra. 
The purpose of this paper is provide such calculation based on the microscopic treatment of 
the electron-magnon interaction in a tunnel ferromagnetic junction.   

Before going to the calculation details given in Secs.~\ref{sec_tunnel}, \ref{sec_spectr} and Appendix \ref{sec_app}, 
in the next section we briefly discuss the approach and main results of the paper.  

\begin{figure}[t]
\begin{center}
\resizebox{0.8\columnwidth}{!}{%
  \includegraphics{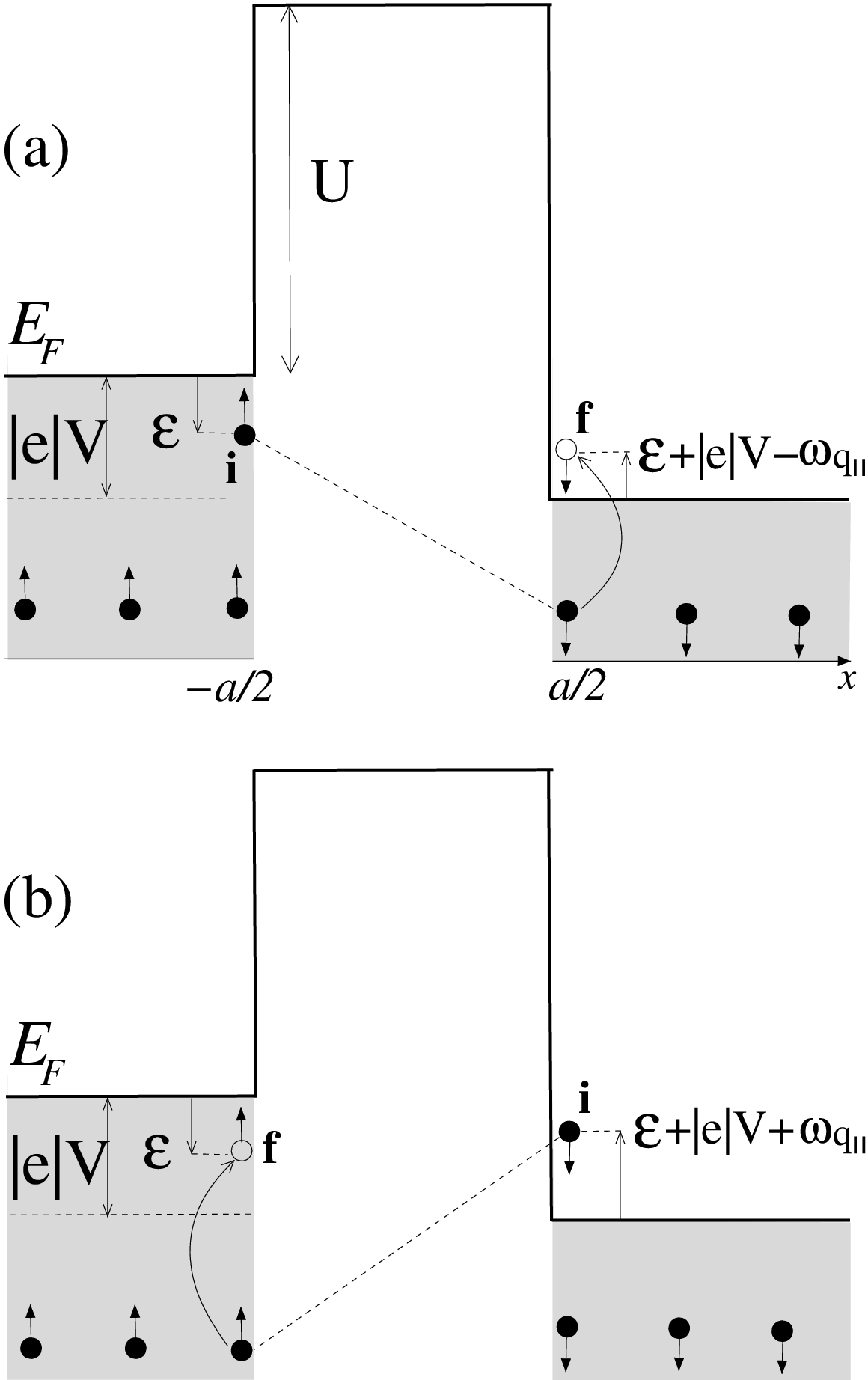}
}
\end{center}
\caption{
Exchange-induced spin-flip tunneling with magnon excitation 
in a junction between half-metallic ferromagnets with antiparallel magnetizations: 
(a) tunneling spin-up electron from the left-hand ferromagnet excites 
via the Coulomb exchange interaction a spin-down electron into a state above the Fermi level in the other ferromagnet. 
This is accompanied by a spin-1 (magnon) excitation of the Fermi sea in the right-hand ferromagnet. 
(b) temperature-stimulated counteracting tunelling process (see also text).
${\bf q}_\|$ and $\omega_{{\bf q}_\|}\leq |e|V$ are the magnon wave-vector and energy,
$a$ and $U$ are the barrier thickness and height, respectively.
}
\label{fig:1}
\end{figure}

\section{Overview of the approach and results}
\label{sec_overview}

The surface-magnon-assisted tunneling is shown schematically in Fig.~1 for a biased junction comprising two fully-polarized (half-metallic) ferromagnets 
with antiparallel (AP) magnetizations. In the process shown in Fig.~1a the magnon emission occurs in the course of the Coulomb exchange interaction between a tunneling spin-up electron from the left-hand ferromagnet and a spin-down electron from the right-hand ferromagnet. The latter is excited into a state above the Fermi level, 
leaving behind a spin-1 excitation of the Fermi sea which in the mean-field approach~\cite{Yosida96} is treated as a spin wave. 
Its energy $\omega_{{\bf q}_\|}$ is limited by the applied bias, i.e. $0 < \omega_{{\bf q}_\|} \leq |e|V$. 
The corresponding inelastic current can be calculated using the nonequilibrium Green's function formalism (e.g. Refs.\cite{Caroli71,Cuevas96} and \cite{GT02a}), 
yielding the following result for the tunneling spectrum:

\begin{eqnarray}
&&
\frac{d^2I^{AP}}{dV^2}=
C^{AP}
\Omega^{-1}
\sum\limits_{{\bf q}_\|}
\frac{{\cal V}^2({\bf q}_\|)}{{\cal V}^2(0)}\int d\epsilon\times
\label{IETS_half}\\
&&
\times 
\left[
n^\prime(\epsilon)n^\prime(\epsilon +|e|V-\omega_{{\bf q}_\|})-
n^\prime(\epsilon +|e|V+\omega_{{\bf q}_\|})n^\prime(\epsilon)
\right],
\nonumber
\end{eqnarray}
Here the junction parameters, such as barrier transparency, 
electron band parameters of the ferromagnets etc., are absorbed in the constant $C^{AP}$ [see Eq. (\ref{C}) in Sec. \ref{sec_spectr}]. 
$\Omega=A/4\pi D$ is the surface MDOS, with $A$ and $D$ being the junction area and the spin stiffness, respectively. 
The matrix element of the electron-magnon coupling is expressed through the Fourier transform
of the Coulomb interaction, ${\cal V}({\bf q}_\|)=2\pi e^2/(\kappa^2 + q_\|^2)^{1/2}$. 
Since the Coulomb interaction occurs across the insulating barrier 
it is assumed weakly screened and, therefore, the electron-magnon coupling in Eq. (\ref{IETS_half}) 
depends on the magnon-wave vector in the junction plane, ${\bf q}_\|$ 
($\kappa^{-1}$ is the screening radius, see Sec. \ref{sec_tunnel} for details). 
In the first term of Eq. (\ref{IETS_half}) the Fermi occupation numbers $n(\epsilon)$ and 
$n(\epsilon +|e|V-\omega_{{\bf q}_\|})$ correspond to the initial (i) 
and final (f) electron states of the process shown in Fig.~1a (the prime denotes the derivative). 
The second term originates from a counteracting exchange process (Fig.~1b) 
where in the initial state a spin-down electron has energy $\epsilon +|e|V+\omega_{{\bf q}_\|}$ 
above the Fermi level in the right-hand system whereas in the final state a spin-up electron has energy $\epsilon$ 
with respect to the Fermi level in the left-hand system. 
For a positive bias voltage $V>0$ this process is only possible at finite temperature $T$ when the occupation number 
of the initial state $n(\epsilon +|e|V+\omega_{{\bf q}_\|})$ is nonzero.

\begin{figure}[t]
\begin{center}
\resizebox{0.8\columnwidth}{!}{%
  \includegraphics{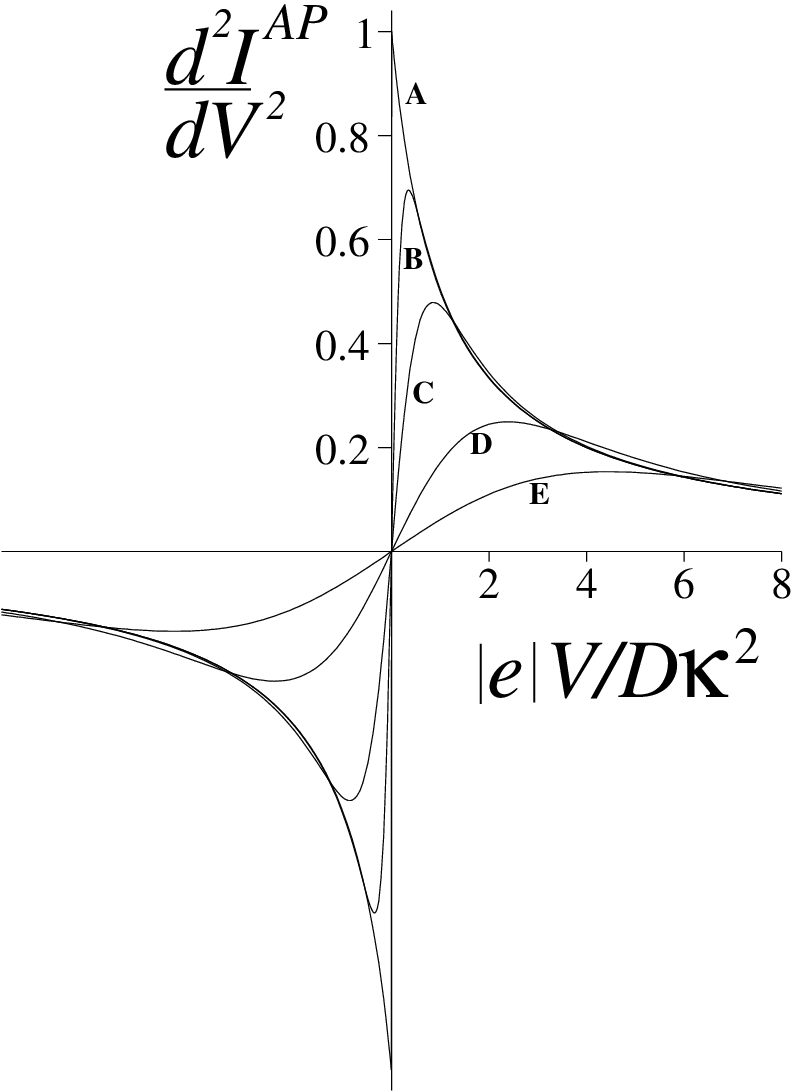}
}
\end{center}
\caption{Second derivative of the tunneling current [in units of constant $C^{AP}$, see Eq. (\ref{IETS_half})] 
for different $k_BT/D\kappa^2$: 
(A)  $0$, (B)  $0.1$, (C)  $0.3$, (D)  $1$, and (E)  $2$,
where $D\kappa^2$ is the characteristic energy of a magnon with wave vector equal to the inverse screening raduis $\kappa$.}
\label{fig:2}
\end{figure}

Figure 2 shows the bias voltage dependence of the derivative $d^2I/dV^2$ (\ref{IETS_half}). 
It vanishes at $V=0$, which is the consequence of the wave-vector dependence 
of the electron-magnon interaction. For decreasing $k_BT\kappa/D\kappa^2$  
the slope of $d^2I/dV^2$ at $V=0$ increases, indicating that the shape of the $d^2I/dV^2$ curves 
is very sensitive to the ${\bf q}_\|$ dependence of ${\cal V}({\bf q}_\|)$ (cf. curves A and E in Fig. 2).  
In the limit $k_BT/D\kappa^2\to 0$ the second derivative is discontinuous at $V=0$ (curve A), 
which is sometimes referred to as the zero-bias anomaly~\cite{Zhang97}.
However, at any finite temperature $T$ the derivative $d^2I/dV^2$ is regular 
and has two antisymmetric peaks at finite bias voltages. 
As we see below this prediction holds for ferromagnets with arbitrary spin-polarization and 
is consistent with several experiments~\cite{Moo98,Jap1,Jap2,Jap3} 
which also reported nonmonotonic inelastic tunneling spectra.

\section{Tunneling Hamiltonian and current operators for interacting electrons}
\label{sec_tunnel}

In order to understand how the matrix elements of the electron-magnon interaction enter the inelastic tunneling spectra, 
the tunneling Hamiltonian and current operators must be derived for interacting electrons in a ferromagnetic junction. 
In this section we derive these operators using the method of effective  boundary conditions. The idea is to solve the equations of motion for the field 
operators of the interacting electrons inside the barrier and "eliminate" this region by expressing the tunneling coupling 
in the form of effective boundary conditions for the "right" and "left" electrons. This can be seen as the continuum version 
of the corresponding recursive Green's function calculations.  

We consider a contact of a large area $A$ between two ferromagnetic metals separated  
by a tunnel barrier (see Fig. 3). The barrier is characterized by thickness $a$ and 
the length of the electron penetration $\lambda=\hbar/(2mU)^{1/2}$ which depends on 
the electron effective mass $m$ and the barrier height $U$ measured with respect to 
the Fermi level. For $U$ of the order of the Fermi energy, $U\sim E_F\gg eV$ ~\cite{Moo98,Moo00}, 
one can neglect the energy and momentum dependence of the electron penetration length. 
The thickness of the barrier $a$ is normally much greater than $\lambda$  
and the density of band electrons inside the barrier, 
$\sim{\rm e}^{-a/\lambda}n_0$ is exponentially reduced compared to that 
in the leads $n_0$, leading to weaker screening of Coulomb electron-electron interactions across
the barrier. 

\begin{figure}[t]
\begin{center}
\resizebox{0.7\columnwidth}{!}{%
  \includegraphics{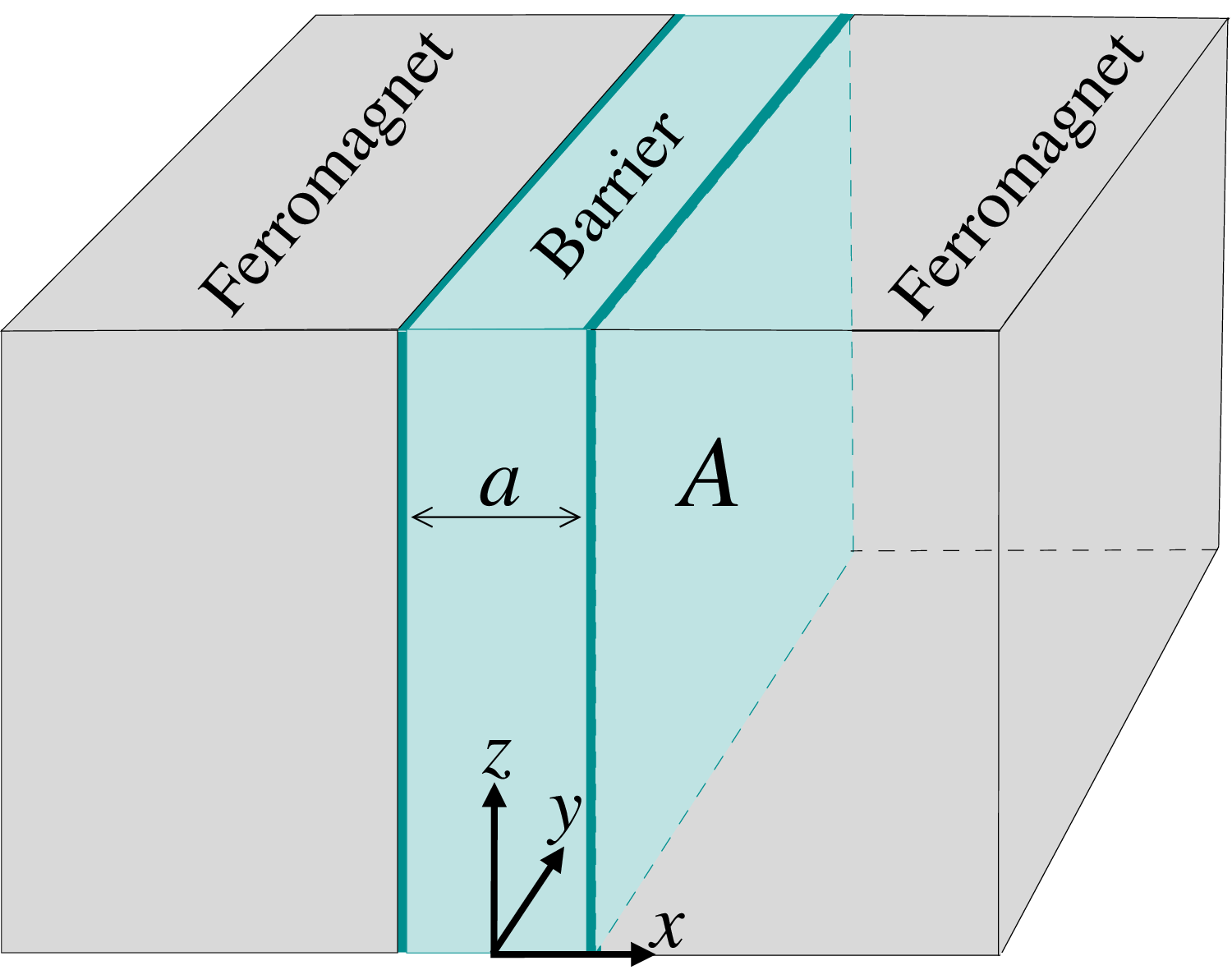}
}
\end{center}
\caption{Schematic view of a ferromagnetic tunnel junction with barrier thickness $a$ and contact area $A$.}
\label{fig:3}
\end{figure}

In order to treat such interactions, we will use the Coulomb potential 
${\cal V}({\bf R})=e^2{\rm e}^{-\kappa R}/R$ with the screening radius 
$\kappa^{-1}\sim 1/({\rm e}^{-a/\lambda}n_0)^{1/3}$ 
much greater than that in the metals ($\sim n_0^{-1/3}$), where 
${\bf R}=\{x;{\bf r}\}$ with the $x$-axis perpendicular to the interface and 
the position vector in the interface plane ${\bf r}$. 
In the case of the experiments described in Refs.~\cite{Moo98,Moo00}, 
$\kappa^{-1}$ can be estimated as being greater than the thickness of the barrier $a$, 
which allows one to neglect the $x$-dependence of the Coulomb potential: 
${\cal V}({\bf r})=e^2{\rm e}^{-\kappa r}/r$.        
Out of all the interaction terms we will only retain those due to  
the exchange of two electrons in the states with opposite spins 
$\alpha$ and $-\alpha$ since they are known to result in the exchange-induced spin
excitations~\cite{Yosida96}. The exchange Hamiltonian can be written as
\begin{eqnarray}
&&
H_{ex}=-\sum\limits_{\alpha,{\bf k}_2\not={\bf k}_1+{\bf q}}
\frac{{\cal V}_{{\bf k}_2-{\bf k}_1-{\bf q}}}{2A}
\int\limits_{-\frac{a}{2}}^{\frac{a}{2}} dx_1dx_2
\times\label{Ex}\\
&&
\times
\chi^\dag_{\alpha{\bf k}_2-{\bf q}}(x_1)
\chi_{-\alpha{\bf k}_2}(x_2)
\chi^\dag_{-\alpha{\bf k}_1+{\bf q}}(x_2)
\chi_{\alpha{\bf k}_1}(x_1),
\nonumber
\end{eqnarray}
where the operators 
$\chi_{\alpha{\bf k}}(x)$ and $\chi^\dag_{\alpha{\bf k}}(x)$ 
annihilate and create, respectively, an electron with spin $\alpha$ and 
wave-vector ${\bf k}$ parallel to the interface at point $x$ inside the barrier 
(the index $\|$ is dropped from now on). The Fourier transform of 
the Coulomb potential is given by
\begin{equation}
{\cal V}_{{\bf q}}=2\pi e^2
\int\limits_0^\infty rdr\frac{ {\rm e}^{-\kappa r} }{r}J_0(rq)=
\frac{2\pi e^2}{(\kappa^2+q^2)^{1/2}},
\label{Vq}
\end{equation}
where $J_0(x)$ is the Bessel function.
The equation of motion for $\chi_{\alpha{\bf k}}(x)$ can be written as 
\begin{eqnarray}
&&
[\partial^2_x-\lambda^{-2}]\chi_{\alpha{\bf k}}(x)=
-\frac{2m}{\hbar^2A}\sum\limits_{{\bf k}_1{\bf q}}{\cal V}_{{\bf k}-{\bf k}_1}
\times\label{Eq}\\
&&
\times
\int\limits_{-\frac{a}{2}}^{\frac{a}{2}} dx_1
\chi_{-\alpha{\bf k}+{\bf q}}(x_1)
\chi^\dag_{-\alpha{\bf k}_1+{\bf q}}(x_1)
\chi_{\alpha{\bf k}_1}(x),
\nonumber
\end{eqnarray}
with usual boundary conditions imposed by the continuity of 
the particle current at the barrier walls:
\begin{eqnarray}
  \chi_{\alpha{\bf k}}(\pm a/2)=\Psi^{r,l}_{\alpha{\bf k}},\quad
  \partial_x\chi_{\alpha{\bf k}}(\pm a/2)=
  \partial_x\Psi^{r,l}_{\alpha{\bf k}},
  \label{bound}
\end{eqnarray}
where $\Psi^{r,l}_{\alpha{\bf k}}\equiv \Psi_{\alpha{\bf k}}(\pm a/2)$ 
are the operators of the right and left systems acting at the barrier boundaries. 
The eigenstates in the left and right systems can at this stage be arbitrary.

To first order in the interaction the solution of Eq. (\ref{Eq}) is 
\begin{eqnarray}
&&
\chi_{\alpha{\bf k}}(x)=\chi^{(0)}_{\alpha{\bf k}}(x)-
\sum\limits_{{\bf k}_1{\bf q}}\frac{{\cal V}_{{\bf k}-{\bf k}_1}}{A}
\int\limits_{-\frac{a}{2}}^{\frac{a}{2}} dx_1dx_2
\times\label{Sol}\\
&&
\times
{\cal G}^{(0)}(x,x_1)
\chi^{(0)}_{-\alpha{\bf k}+{\bf q}}(x_2)
\chi^{(0)\dag}_{-\alpha{\bf k}_1+{\bf q}}(x_2)
\chi^{(0)}_{\alpha{\bf k}_1}(x_1),
\nonumber
\end{eqnarray} 
where the operators of the non-interactive system
\begin{eqnarray}
&&
\chi^{(0)}_{\alpha{\bf k}}(x)=\phi^r(x)\Psi^r_{\alpha{\bf k}}+
\phi^l(x)\Psi^l_{\alpha{\bf k}},
\label{Sol0}\\
&&
\phi^{r,l}(x)=\frac{\sinh(a/2 \pm x)/\lambda}{\sinh(a/\lambda)},
\label{rl}
\end{eqnarray}
are linear combinations of the two fundamental solutions (\ref{rl}) of unperturbed 
equation (\ref{Eq}). ${\cal G}^{(0)}$ is the Green function 
of the unperturbed equation which can be constructed using $\phi^{r,l}(x)$ as follows: 
\begin{eqnarray}
&
{\cal G}^{(0)}(x_1,x_2)=-2m\hbar^{-2}\lambda\sinh(a/\lambda)
\times&
\label{G0}\\
&
\times
\left\{
\begin{array}{cc}
\phi^l(x_1)\phi^r(x_2) & x_1\geq x_2,\\
\phi^l(x_2)\phi^r(x_1) & x_2\geq x_1,
\end{array}
\right.&
\nonumber
\end{eqnarray}
Note that our choice of the integration constants ensures that solution (\ref{Sol}) 
matches the operators of the right $\Psi^r_{\alpha{\bf k}}$ 
and left $\Psi^l_{\alpha{\bf k}}$ systems at the boundaries $x=a/2$ and $x=-a/2$ 
(first boundary condition in Eq. (\ref{bound})).  

Solution (\ref{Sol}) is expressed in terms of $\Psi^r_{\alpha{\bf k}}$ 
and $\Psi^l_{\alpha{\bf k}}$ and contains no more constants to be determined. 
Inserting it into the boundary conditions for the derivatives (\ref{bound}) and 
evaluating the integrals over the coordinates, we find 
\begin{eqnarray}
&&
  \partial_x\Psi^r_{\alpha{\bf k}}=
  \frac{\Psi^r_{\alpha{\bf k}}-2{\rm e}^{-a/\lambda}\Psi^l_{\alpha{\bf k}}}{\lambda} 
  -
  {\rm e}^{-a/\lambda}
  \sum\limits_{{\bf k}_1{\bf q}}
  \frac{\tau {\cal V}_{{\bf k}-{\bf k}_1}}{A\hbar}\times
  \nonumber\\
&&
  \times
  \left(
  \Psi^l_{-\alpha{\bf k}+{\bf q}}
  \Psi^{l\dag}_{-\alpha{\bf k}_1+{\bf q}}+
  \Psi^r_{-\alpha{\bf k}+{\bf q}}
  \Psi^{r\dag}_{-\alpha{\bf k}_1+{\bf q}}
  \right)
  \Psi^l_{\alpha{\bf k}_1},\quad
  \label{boundR}
\end{eqnarray}
\begin{eqnarray}
&&
  \partial_x\Psi^l_{\alpha{\bf k}}=
  \frac{-\Psi^l_{\alpha{\bf k}}+2{\rm e}^{-a/\lambda}\Psi^r_{\alpha{\bf k}}}{\lambda} 
  +
  {\rm e}^{-a/\lambda}
  \sum\limits_{{\bf k}_1{\bf q}}
  \frac{\tau {\cal V}_{{\bf k}-{\bf k}_1}}{A\hbar}\times
  \nonumber\\
&&
  \times
  \left(
  \Psi^l_{-\alpha{\bf k}+{\bf q}}
  \Psi^{l\dag}_{-\alpha{\bf k}_1+{\bf q}}+
  \Psi^r_{-\alpha{\bf k}+{\bf q}}
  \Psi^{r\dag}_{-\alpha{\bf k}_1+{\bf q}}
  \right)
  \Psi^r_{\alpha{\bf k}_1},\quad
  \label{boundL}
\end{eqnarray}
\begin{eqnarray}
  \tau=ma\lambda/\hbar=a(m/2U)^{1/2}.
  \label{time} 
\end{eqnarray}
These equations now serve as {\em effective boundary conditions} for the right and left systems. 
Because of the tunneling, the right and left operators are mixed in Eqs. (\ref{boundR}) and (\ref{boundL}). 
In the limit ${\rm e}^{-a/\lambda}\to 0$, the coupling vanishes:
\begin{eqnarray}
\partial_x\Psi^r_{\alpha{\bf k}}=
\frac{\Psi^r_{\alpha{\bf k}}}{\lambda}, \quad
\partial_x\Psi^l_{\alpha{\bf k}}=
-  \frac{\Psi^l_{\alpha{\bf k}}}{\lambda},
\label{bound_is}
\end{eqnarray} 
and the boundary conditions (\ref{bound_is}) describe isolated right and left systems 
with the particle current vanishing at both $x=a/2$ and $x=-a/2$. Boundary
conditions (\ref{bound_is}) will be used in Appendix \ref{sec_app} to introduce the eigenstates in the isolated right and left systems. 
When evaluating the integrals over the coordinates we have only taken into account terms linear in ${\rm e}^{-a/\lambda}\ll 1$ in Eqs. (\ref{boundR}) and (\ref{boundL}). 

Note that the exchange interaction results in the mixing of the left and right operators with opposite spins in the boundary conditions (\ref{boundR}) and (\ref{boundL}), 
which takes into account inelastic spin-flip processes during the tunneling. These terms are proportional 
to the tunneling time $\tau$ (\ref{time}) assumed to be sufficiently short to justify 
the use of perturbation theory. 

We now use the effective boundary conditions (\ref{boundR}) and (\ref{boundL}) to derive 
the microscopic tunneling current operator and transfer Hamiltonian.
Both the current operator and the transfer Hamiltonian will be expressed in terms of the field 
operators taken at the left, $\Psi^l_{\alpha{\bf k}}$ and right, $\Psi^r_{\alpha{\bf k}}$ 
boundaries of the barrier. As earlier, no particular eigenstates in the left and right systems will 
be assumed during the derivation procedure.  

As the boundary conditions (\ref{boundR}) and (\ref{boundL}) conserve the current density, 
the total current operator ${\hat I}$ can be related to the current density operator at any of the boundaries, 
e.g. at the left one:  
${\hat I}=\frac{ie\hbar}{2m}
\sum_{\alpha{\bf k}}(\partial_x\Psi^{l\dag}_{\alpha{\bf k}}\Psi^l_{\alpha{\bf k}}-h.c.)$
with the derivative $\partial_x\Psi^{l\dag}_{\alpha{\bf k}}$ given by boundary condition (\ref{boundL}). 
The current can be written as the sum of an elastic and an inelastic contributions: 
\begin{equation}
{\hat I}={\hat I}_{el}+{\hat I}_{in}, 
\label{I_tot}
\end{equation}
where
\begin{eqnarray}
  {\hat I}_{el}
  &=&
  \frac{ie{\cal T}}{\hbar}
  \sum_{\alpha{\bf k}}(\Psi^{r\dag}_{\alpha{\bf k}}\Psi^l_{\alpha{\bf k}}-h.c.),\quad
  {\cal T}=\frac{\hbar^2 {\rm e}^{-a/\lambda}}{m\lambda},\quad
  \label{I_el}\\
  {\hat I}_{in}
  &=&
  \frac{ie{\cal T}}{\hbar}
  \sum\limits_{\alpha,{\bf k}_2\not={\bf k}_1+{\bf q}}
  \frac{\tau{\cal V}_{{\bf k}_2-{\bf k}_1-{\bf q}}}{A\hbar}
  \times\label{I_in}\\
  &\times&
  \frac{\lambda}{2}
  \left( 
  \Psi^{r\dag}_{\alpha{\bf k}_2-{\bf q}}
  \Psi^r_{-\alpha{\bf k}_2}
  \Psi^{r\dag}_{-\alpha{\bf k}_1+{\bf q}}
  \Psi^l_{\alpha{\bf k}_1}  +\right.
\nonumber\\
  &+&\left.
  \Psi^{r\dag}_{\alpha{\bf k}_1-{\bf q}}
  \Psi^l_{-\alpha{\bf k}_1}
  \Psi^{l\dag}_{-\alpha{\bf k}_2+{\bf q}}
  \Psi^l_{\alpha{\bf k}_2}
  -h.c.\right).
  \nonumber
\end{eqnarray}
In these equations ${\cal T}$ plays the role of a one-particle "hopping" parameter between 
the left and the right systems.

Note that for electrons with close enough wave-vectors
\begin{eqnarray}
|{\bf k}_2-{\bf k}_1|\ll (\kappa^2+q^2)^{1/2},
\label{states}
\end{eqnarray}
the matrix elements of the interaction in Eq. (\ref{I_in}) 
are independent of both ${\bf k}_2$ and ${\bf k}_1$: 
${\cal V}_{{\bf k}_2-{\bf k}_1-{\bf q}}\approx {\cal V}_{{\bf q}}$. 
In this case the sum over ${\bf k}_2$ picks up the products of 
the operators describing simultaneous creation of a hole and an electron 
with opposite spins. The superpositions of such electron-hole pair operators 
are related to {\it electron spin operators}. 
Let us introduce first the operator of the total spin of the electrons 
penetrating into the barrier from the right ferromagnet:  
\begin{eqnarray}
&&
 S^r_z=\frac{1}{2}
 \sum\limits_{\bf k}\int\limits_0^{a/2} dx
 [
 \chi^{(0)\dag}_{\uparrow{\bf k}}(x)\chi^{(0)}_{\uparrow{\bf k}}(x)-
 \chi^{(0)\dag}_{\downarrow{\bf k}}(x)\chi^{(0)}_{\downarrow{\bf k}}(x)]\approx
\nonumber\\
&&
 \approx\frac{1}{2}
 \sum\limits_{\bf k}\frac{\lambda}{2}
 [
 \Psi^{r\dag}_{\uparrow{\bf k}}\Psi^r_{\uparrow{\bf k}}-
 \Psi^{r\dag}_{\downarrow{\bf k}}\Psi^r_{\downarrow{\bf k}}].
\label{Sz}
\end{eqnarray}
Here $\chi^{(0)}_{\alpha{\bf k}}(x)$ (\ref{Sol0}) 
exponentially decays over the distances of the order of $\lambda$ from the right boundary. 
In Appendix \ref{sec_app} we calculate $S^r_z$ [see, Eq. (\ref{Sz1})] and show that it is  
a macroscopic quantity much greater than unity, so that it can be treated as a 
classical spin. Then the operators $S^{r+}$ and $S^{r-}$, 
raising and lowering the total spin $S^r_z$, can be introduced as 
\begin{eqnarray}
&&
 S^{r+}_{\bf q}=\frac{1}{(2|S_z|)^{1/2}}
 \sum\limits_{\bf k}\int\limits_0^{a/2} dx
 \chi^{(0)\dag}_{\uparrow{\bf k}-{\bf q}}(x)
 \chi^{(0)}_{\downarrow{\bf k}}(x)\approx
\nonumber\\
&&
 \approx 
 \frac{1}{(2|S_z|)^{1/2}}
 \sum\limits_{\bf k}\frac{\lambda}{2}
 \Psi^{r\dag}_{\uparrow{\bf k}-{\bf q}}
 \Psi^{r}_{\downarrow{\bf k}},
\label{S+}\\
&&
 S^{r-}_{\bf q}=\frac{1}{(2|S_z|)^{1/2}}
 \sum\limits_{\bf k}\int\limits_0^{a/2} dx
 \chi^{(0)\dag}_{\downarrow{\bf k}-{\bf q}}(x)
 \chi^{(0)}_{\uparrow{\bf k}}(x)\approx
\nonumber\\
&&
 \approx 
 \frac{1}{(2|S_z|)^{1/2}}
 \sum\limits_{\bf k}\frac{\lambda}{2}
 \Psi^{r\dag}_{\downarrow{\bf k}-{\bf q}}
 \Psi^{r}_{\uparrow{\bf k}}.
 \label{S-}
\end{eqnarray} 
They are normalized in the usual way to satisfy the boson commutation
relation: $S^{r+}_{\bf q}S^{r-}_{-\bf q}-S^{r-}_{-\bf q}S^{r+}_{\bf q}=sign(S^r_z)$. 
Thus, $S^{r+}_{\bf q}$ and $S^{r-}_{-\bf q}$ are the magnon annihilation 
and creation operators, respectively, for positive $S^r_z$ and vice versa 
for negative $S^r_z$. Also, since the operators 
$S^{r,l+}_{\bf q}$ and $S^{r,l-}_{\bf q}$ change the electron spin in the 
surface layers of thickness $\lambda$, 
they describe {\it surface} magnons. In what follows we adopt the parabolic magnon dispersion 
$\omega^{r,l}_{\bf q}=D^{r,l}q^2$, where $D^{r,l}$ is the spin stiffness. 

Keeping in equation (\ref{I_in}) only the coherent terms with wave-vectors satisfying Eq. 
(\ref{states}), one can express the inelastic current in terms of the 
magnon operators (\ref{S+}) and (\ref{S-}) as    
\begin{eqnarray}
 &&
  {\hat I}_{in}=(2|S_z|)^{1/2}\frac{ie {\cal T}}{\hbar}
  \sum\limits_{{\bf k}{\bf q}}
  \frac{\tau {\cal V}_{\bf q}}{A\hbar}
  \left[
  (S^{r+}_{\bf q}+S^{l+}_{\bf q})
  \Psi^{r\dag}_{\downarrow{\bf k}+{\bf q}}
  \Psi^l_{\uparrow{\bf k}}+
  \right.  
  \nonumber\\
 &&
  \left. 
 +(S^{r-}_{\bf q}+S^{l-}_{\bf q})
  \Psi^{r\dag}_{\uparrow{\bf k}+{\bf q}}
  \Psi^l_{\downarrow{\bf k}}
  -h.c.\right],
  \label{I_spin}
\end{eqnarray}
where for simplicity $|S^r_z|=|S^l_z|\equiv |S_z|$. This equation along with Eq. (\ref{I_el}) 
define the total tunneling current operator (\ref{I_tot}). To introduce the tunneling Hamiltonian 
we write the tunneling current as the rate of change of the particle number 
in one of the systems (e.g in the left one): 
\begin{eqnarray}
&
{\hat I}=-e{\dot {\hat N}}_L=\frac{ie}{\hbar}[{\hat N}_L,H_T],&
\label{I_HT}\\
&
{\hat N}_L=\sum_{\alpha{\bf k}}\int\limits_{x\leq -a/2}
\Psi^{\dag}_{\alpha{\bf k}}(x)\Psi_{\alpha{\bf k}}(x)dx,&
\nonumber
\end{eqnarray}
which involves the commutator of the particle number operator in the left system, ${\hat N}_L$, and 
the tunneling Hamiltonian $H_T$. It is straightforward to verify that the tunneling Hamiltonian of the form  
\begin{eqnarray}
  H_T=
  &-&
  {\cal T}
  \sum_{\alpha{\bf k}}(\Psi^{r\dag}_{\alpha{\bf k}}\Psi^l_{\alpha{\bf k}}+h.c.)-
  \label{H_T}\\
  &-&
  (2|S_z|)^{1/2}{\cal T}
  \sum\limits_{{\bf k}{\bf q}}
  \frac{\tau {\cal V}_{\bf q}}{A\hbar}
  \left[
  (S^{r+}_{\bf q}+S^{l+}_{\bf q})
  \Psi^{r\dag}_{\downarrow{\bf k}+{\bf q}}
  \Psi^l_{\uparrow{\bf k}}+
  \right.
  \nonumber\\
  &&
  +\left.
  (S^{r-}_{\bf q}+S^{l-}_{\bf q})
  \Psi^{r\dag}_{\uparrow{\bf k}+{\bf q}}
  \Psi^l_{\downarrow{\bf k}}
  +h.c.\right],
  \nonumber
\end{eqnarray}
satisfies Eq. (\ref{I_HT}) with ${\hat I}={\hat I}_{el}+{\hat I}_{in}$ given by Eqs. (\ref{I_el}) and (\ref{I_spin}). 
     
The first term in Eq. (\ref{H_T}) is the real-space version of the well-known elastic tunneling Hamiltonian (see e.g. Refs.~\cite{Cohen62,Duke69}). 
Expanding the operators 
$\Psi^{r,l}_{\alpha{\bf k}}\equiv \Psi_{\alpha{\bf k}}(\pm a/2)$ in the eigenstates of the isolated 
right and left systems, one can go over to the momentum representation used in Ref.~\cite{Cohen62}. 
The advantage of the coordinate representation (\ref{H_T}) is that the tunneling matrix element 
${\cal T}$ (\ref{I_el}) is a constant, which makes perturbation theory in the coordinate representation simpler.
The second term in Eq. (\ref{H_T}) describes inelastic electron tunneling accompanied by the emission (absorption) of a surface magnon. 
It is similar to that used in Ref.~\cite{Zhang97}. However, in the present model the electron-magnon coupling in the Hamiltonian $H_T$ (\ref{H_T}) 
comes from the exchange interaction of the itinerant electrons mediated by the Coulomb potential and is characterized by the matrix element 
${\cal V}_{\bf q}$ (\ref{Vq}) which depends on the magnon wave-vector ${\bf q}$ and, therefore, on its energy.

\section{Elastic and inelastic contributions to the tunneling current}
\label{sec_spectr}

To calculate the current-voltage characteristics $I(V)$ one should perform the statistical averaging of the tunneling current operator $\hat{I}$ 
[Eqs. (\ref{I_tot}), (\ref{I_el}) and (\ref{I_spin})] over the nonequilibrium state with finite difference $eV$ 
in chemical potentials of the left and right ferromagnets. 
Since the current operator $\hat{I}$ is linear in tunneling matrix element ${\cal T}$, 
it is sufficient to use the first order perturbation theory with respect to the tunneling Hamiltonian $H_T\propto {\cal T}$ (\ref{H_T}), 
which yields the lowest order result $\propto {\cal T}^2$. 
Leaving aside these standard calculations (involving the nonequilibrium electron and magnon Green's functions, see e.g. Appendix of Ref. \cite{GT02a}), 
we proceed to the analysis of the tunneling current-voltage characteristics $I(V)$ and 
first briefly discuss the elastic contribution $I_{el}$ for the parralel (P) and antiparallel (AP) 
magnetization orientations: 
\begin{eqnarray}
&&
I^P_{el}=\frac{2\pi e^2V{\cal T}^2}{\hbar} (\rho_{MM}+\rho_{mm}),
\nonumber\\
&&
I^{AP}_{el}=\frac{2\pi e^2V{\cal T}^2}{\hbar}(\rho_{mM}+\rho_{Mm}),
\label{elastic}\\
&&
\rho_{MM}=\sum\limits_{{\bf k}}\rho_{lM}(E_F,{\bf k})\rho_{rM}(E_F,{\bf k}),
\label{MM}\\
&&
\rho_{mm}=\sum\limits_{{\bf k}}\rho_{lm}(E_F,{\bf k})\rho_{rm}(E_F,{\bf k}),
\label{mm}\\
&&
\rho_{mM}=\sum\limits_{{\bf k}}\rho_{lm}(E_F,{\bf k})\rho_{rM}(E_F,{\bf k}),
\label{mM}\\
&&
\rho_{Mm}=\sum\limits_{{\bf k}}\rho_{lM}(E_F,{\bf k})\rho_{rm}(E_F,{\bf k}).
\label{Mm}
\end{eqnarray}
Here $\rho_{r,lM}(E_F,{\bf k})$ and $\rho_{r,lm}(E_F,{\bf k})$ are the majority (M) and minority
(m) local electron spectral densities related to the retarded and advanced  
electron Green functions ${\cal G}^{R,A}_{r,l}(E_F,{\bf k})$ at the boundaries of the ferromagnets:
\begin{eqnarray}
&&
\rho_{r,l}(E_F,{\bf k})
=
\frac{{\cal G}^A_{r,l}(E_F,{\bf k})-
{\cal G}^R_{r,l}(E_F,{\bf k})}{2\pi i}.
\quad
\label{rho}
\end{eqnarray}
They are taken at the Fermi energy for $|eV|\ll E_F$. 
As in our case the parallel wave-vector ${\bf k}$ is conserved upon the tunneling (coherent tunneling), 
the current (\ref{elastic}) is proportional to the trace of the product of two spectral 
densities. For the parallel alignmet it is proportional to $\rho_{MM}+\rho_{mm}$
since the tunneling occurs independently between the majority and minority bands 
whereas for the antiparallel case carriers tunnel from majority to minority bands (and vice versa) 
and hence $I_{el}\propto \rho_{mM}+\rho_{Mm}$. The degree of spin-polarization 
${\cal P}=(I^{P}_{el}-I^{AP}_{el})/I^{P}_{el}$ of the elastic current 
is given by
\begin{eqnarray}
{\cal P}=
\frac{\rho_{MM}+\rho_{mm}-\rho_{mM}-\rho_{Mm}}{\rho_{MM}+\rho_{mm}}<1.
\label{P} 
\end{eqnarray} 
For incoherent tunneling where the parallel momentum is not conserved, 
${\cal P}$ would be expressed in terms of the local densities of the states 
$\sum_{{\bf k}}\rho_{r,lM}(E_F,{\bf k})$
and $\sum_{{\bf k}}\rho_{r,lm}(E_F,{\bf k})$ rather than 
the momentum convolutions of the spectral densities (\ref{MM})--(\ref{Mm}). In Appendix \ref{sec_app} 
we give the expressions for the traces (\ref{MM})--(\ref{Mm}) in terms of the band electron parameters of the ferromagnets 
[see, Eqs. (\ref{MM1}) and (\ref{Mm1})].      

As to the inelastic current $I_{in}$, let us first discuss the antiparallel alignment of the magnetic moments 
for which one finds
\begin{eqnarray}
&&
I^{AP}_{in}=\frac{2\pi |S_ze| {\cal T}^2}{\hbar}
\sum\limits_{{\bf q}}
\left(\frac{\tau {\cal V}_{\bf q}}{A\hbar}\right)^2\int d\epsilon d\omega\times
\label{I_AP}\\
&&
\times
\left[
(\rho_{MM}\Omega_r(\omega,{\bf q})+\rho_{mm}\Omega_l(\omega,{\bf q}))
\right.
\nonumber\\
&&
\times
\left\{
n(\epsilon)[1-n(\epsilon +|e|V-\omega)][1+N(\omega)]
\right.
\nonumber\\
&&
\left.
-[1-n(\epsilon)]n(\epsilon +|e|V-\omega)N(\omega)
\right\}
\nonumber\\
&&
+(\rho_{MM}\Omega_l(\omega,{\bf q})+\rho_{mm}\Omega_r(\omega,{\bf q}))
\nonumber\\
&&
\times
\left\{
n(\epsilon)[1-n(\epsilon +|e|V+\omega)]N(\omega)
\right.
\nonumber\\
&&
\left.\left.
-[1-n(\epsilon)]n(\epsilon +|e|V+\omega)[1+N(\omega)]
\right\}
\right].
\nonumber
\end{eqnarray}
Here we assume that the majority electrons in the left and the right systems are spin-up 
($\uparrow$) and spin-down ($\downarrow$) ones, respectively (Fig.~4). The magnon spectral 
densities $\Omega_{r,l}(\omega,{\bf q})$ are expressed in terms of the advanced and retarded magnon Green functions ${\cal D}^{R,A}_{r,l}(\omega,{\bf q})$ as 
\begin{eqnarray}
\Omega_{r,l}(\omega,{\bf q})
=
\frac{{\cal D}^A_{r,l}(\omega,{\bf q})-
{\cal D}^R_{r,l}(\omega,{\bf q})}{2\pi i}=
\delta(\omega -\omega^{r,l}_{\bf q}).\qquad
\label{Omega}
\end{eqnarray}
The products of the electron, $n(\epsilon)$, and 
magnon, $N(\omega)$, occupation numbers in Eq. (\ref{I_AP}) correspond to various emission and absoption processes.   
For arbitrary spin polarizations of the ferromagnets there are four magnon-emission 
and four magnon-absoption processes. The latter are only possible at finite temperatures when $N(\omega)\not =0$ and generate current 
in the opposite direction with respect to the magnon-emission current. Below we discuss the emission processes, 
shown schematically for $T=0$ in Fig.~4. The details of the absorption processes can be analyzed in the same way.

The processes in Fig.~4a and 4b correspond to the first term in the square brackets in Eq. (\ref{I_AP}) which determines the current 
at positive voltages, when the Fermi level in the left system is higher than that in the right one. 
The process in Fig.~4a has been already discussed in Sec. \ref{sec_overview} for the case 
of half-metallic ferromagnets (Fig.~1). As both initial and final electron states belong to the majority bands, 
the corresponding contribution to the current is 
proportional to $\rho_{MM}$ (\ref{MM}) and the magnon spectral density 
at the right side of the junction $\Omega_r$ (\ref{Omega}). Note that the similar interaction of 
the minority electrons cannot give rise to the magnon emission because it would 
result in the magnetic moment of the right system bigger than that in the ground state.   
However the minority magnon-assisted transport [proportional to $\rho_{mm}$ (\ref{mm})] 
can be realised in a way shown in Fig.~4b: In the course of the exchange interaction, a minority (spin-down) electron from the left side 
excites a majority (spin-up) electron from the same side above the Fermi level in the right system. The latter occupies an empty state 
in the spin-up (minority) conduction band in the right ferromagnet, while a spin-wave excitation of the majority (spin-up) Fermi sea 
is created on the left side of the junction. 
The corresponding contribution to the current (\ref{I_AP}) is proportional 
to the magnon spectral density at the left boundary $\Omega_l$ (\ref{Omega}). 
The processes in Fig.~4c and 4d are generated in negatively biased junctions and described by 
the second term in the square brackets in Eq. (\ref{I_AP}). Although they look similar to 
those shown in Fig.~4a and 4b, in general there is no symmetry because 
the magnon spectral densities at the left and right boundaries need not to be identical: 
$\Omega_r\not =\Omega_l$.   
 
\begin{figure}[t]
\begin{center}
\resizebox{0.7\columnwidth}{!}{%
  \includegraphics{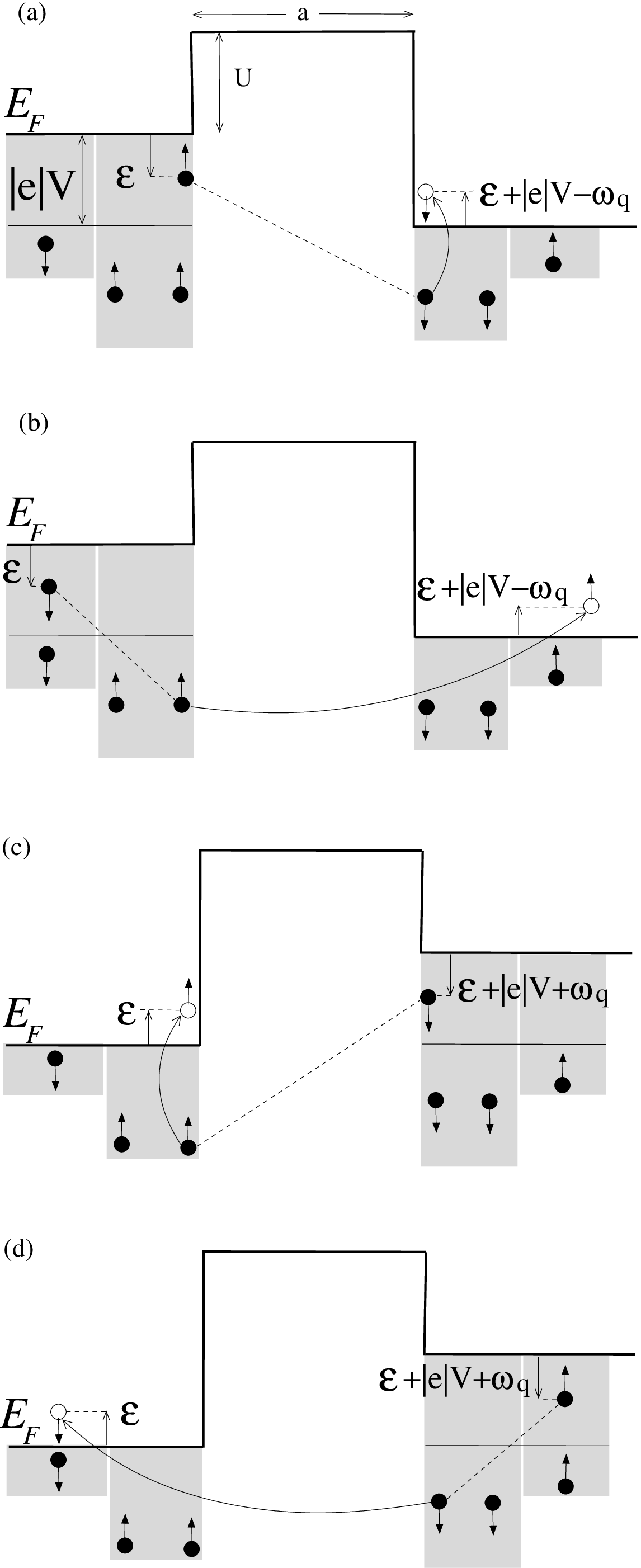}
}
\end{center}
\caption{Exchange-induced spin-flip tunneling with magnon emission between ferromagnets of arbitrary spin-polarization
for antiparallel configuration at zero temperature: 
(a) and (b) assisted tunneling between the majority and minority bands, respectively, 
for positive bias voltage $V>0$, 
(c) and (d) same for a negatively biased junction $V<0$.}
\label{fig:4}
\end{figure}

For the parallel alignment of the magnetic moments, 
when the majority electrons in both ferromagnets are spin-up ones, one can obtain the following expression for the inelastic
current:

\begin{eqnarray}
&&
I^P_{in}=\frac{2\pi |S_ze| {\cal T}^2}{\hbar}
\sum\limits_{{\bf q}}
\left(\frac{\tau {\cal V}_{\bf q}}{A\hbar}\right)^2\int d\epsilon d\omega
\times\label{I_P}\\
&&
\times
(\Omega_r(\omega,{\bf q})+\Omega_l(\omega,{\bf q}))  
\nonumber\\
&&
\times
\left[
\rho_{mM}\,
\left\{
n(\epsilon)[1-n(\epsilon +|e|V-\omega)][1+N(\omega)]
\right.
\right.
\nonumber\\
&&
\left.
-[1-n(\epsilon)]n(\epsilon +|e|V-\omega)N(\omega)
\right\}
\nonumber\\
&&
+
\rho_{Mm}
\,
\left\{
n(\epsilon)[1-n(\epsilon +|e|V+\omega)]N(\omega)
\right.
\nonumber\\
&&
\left.
\left.
-[1-n(\epsilon)]n(\epsilon +|e|V+\omega)[1+N(\omega)]
\right\}
\right].
\nonumber
\end{eqnarray}
Unlike the antiparallel case [see, Eq. (\ref{I_AP})], here the magnon-assisted 
transport is due to the exchange between a minority electron and a majority one. The latter can 
be either from the opposite or the same side of the junction, which explains why 
the current (\ref{I_P}) contains the sum of the magnon spectral densities,
$\Omega_r(\omega,{\bf q})+\Omega_l(\omega,{\bf q})$,  and
is proportional to the traces in the momentum space of the minority and the majority electron spectral densities [see Eqs. (\ref{mM}) and (\ref{Mm})]. 

The nonequilibrium tunneling spectra are given by the second derivatives $d^2I^{AP,P}/dV^2$ 
which are nonzero for the inelastic currents (\ref{I_AP}) and (\ref{I_P}) and vanish for the elastic (linear in $V$) contributions (\ref{elastic}).    
In what follows we present the tunneling spectra for a simpler case of identical ferromagnets. 
When evaluating the second derivatives of the inelastic currents, one finds that 
the equilibrium magnon occupation numbers $N(\omega)$ drop out of the expression 
for $d^2I^{AP,P}/dV^2$, i.e. only the non-equilibrium processes of magnon-emission contribute to the tunneling spectrum. 
The expression for $d^2I^{AP,P}/dV^2$ has the same structure as Eq. (\ref{IETS_half}) and contains coefficients $C^{AP,P}$ given by  
\begin{eqnarray}
C^{AP}=
\frac{2\pi |S_ze^3|\Omega}{\hbar}
\left(\frac{\tau {\cal V}(0)}{A\hbar}\right)^2
{\cal T}^2(\rho_{MM}+\rho_{mm}),
\label{C}
\end{eqnarray}
\begin{eqnarray}
\quad
\frac{
C^{P}
}{
C^{AP}
}=1-{\cal P}.
\label{ratio}
\end{eqnarray}
The last equation is quite interesting since it implies that for identical ferromagnets 
the inelastic spectra for the paralell and antiparallel cases are retaled to each other 
via the degree of spin-polarization (\ref{P}). This can be used for an independent measurement of the spin-polarization of the tunneling current. 
At the same time, the shape of the inelastic tunneling spectrum does not depend on the relative alignment of their magnetic moments and 
the degree of spin-polarization (see, e.g. Fig.~2 for half-metallic ferromagnets). 

In Fig.~2 both excitation energy $|e|V$ and $k_BT$ are normalized by the characteristic energy 
of a surface magnon, $D\kappa^2$,  with the wave-vector equal to the inverse screening radius $\kappa$. 
For $k_BT/D\kappa^2\to 0$ the second derivative is discontinuous at $V=0$ (curve A) recovering the 
zero-bias anomaly due to the emission of surface magnons studied 
theoretically in Ref.~\cite{Zhang97}. As $|e|V$ increases, the wave-vector of the excited
magnon, $\sim\sqrt{|eV|/D }$, becomes larger and for $\sqrt{ |eV|/D }\sim \kappa$ 
the electron-magnon coupling in Eq. (\ref{IETS_half}) becomes strongly 
energy-dependent, leading to a $1/|e|V$ decrease in the tunneling spectrum. 
Finite temperatures (curves B-E) result in the smearing of the zero-bias anomaly 
due to the counter spin-flip processes (Fig.~1b) which lead to a finite-slope increase 
in the response at small $|e|V$ and hence to the formation of {\it two antisymmetric peaks}. 
In agreement with the experimental data of Refs.~\cite{Moo98,Moo00,Jap1,Jap2,Jap3},
at relatively low temperatures ($k_BT < D\kappa^2$, curves B and C) the peaks are sharp. 
As the temperature increases, they broaden and shift towards higher excitation energies 
(curves D and E). At large $|e|V$ all the curves merge showing a temperature-independent behaviour, 
also clearly seen in the experiments~\cite{Moo98,Moo00}. 
It should be noted that for screening radius $\kappa^{-1}\geq a$ the characteristic magnon energy $D\kappa^2$ 
is still much smaller than $k_BT_C$. 

To compare our results with the experimental data of Refs.~\cite{Moo98,Moo00,Jap1,Jap2,Jap3}, 
we estimate the voltage corresponding to the peak positions at low temperatures as $V_P\sim D\kappa^2/|e|$. 
For $\kappa^{-1}\sim a\sim 10\AA$ and spin stiffness typical for transition metals, 
$D\approx 300-500\, meV\times\AA^2$, one obtains $V_P\sim 3-5\, mV$. In Refs.~\cite{Jap2},~\cite{Jap1} and~\cite{Moo98} 
the peaks were observed at $2\, mV$, $12\, mV$ and $17\, mV$, respectively. 
At the same time, the Curie temperature of the ferromagnets corresponds to the voltage of order of $100\, mV$. 
The relation (\ref{ratio}) between the spin-polarization of the current and the peak intensities for the parallel and antiparalell configurations 
is also consistent with experimental data~\cite{Jap1,Jap2}.

The author thanks K.~Richter, J.~Siewert and D.~Weiss for discussions.
The work was supported by the DFG within SFB 689.

\appendix
\section{Local electron spectral densities and spin polarization}
\label{sec_app}

In this appendix we calculate the local electron spectral densities 
$\rho_{MM}$, $\rho_{mm}$ and $\rho_{Mm}$ (\ref{MM})--(\ref{Mm}) and spin polarization $|S_z|$ (\ref{Sz}) which enter 
the inelastic tunneling spectrum via coefficients $C^{AP,P}$ (\ref{C}).  
This will be done for the isolated 
right and left systems which are described by the boundary conditions 
(\ref{bound_is}) where $\lambda$ means the electron penetration length into an infinitely thick 
barrier ($a\to\infty$) of the finite height $U$. We will assume identical ferromagnets
where it is enough to calculate the local spectral densities at the boundary of one of the electrodes, say, the right one $x=a/2$. 
Using the expression for the local electron spectral density in terms of the advanced and retarded Green functions (\ref{rho}), 
one can write
\begin{eqnarray}
&
\rho_{rM,m}(E_F,{\bf k})=\sum_{k_x}\phi^2_{k_x}\left(\frac{a}{2}\right)
\delta(E_F-E_{M,m}(k_x,{\bf k}))=&
\nonumber\\
&
=\lambda^2\sum_{k_x}(\partial_x\phi_{k_x}\left(\frac{a}{2})\right)^2
\delta(E_F-E_{M,m}(k_x,{\bf k})),&
\label{rho1}
\end{eqnarray}  
where we introduce the eigenstates $\phi_{k_x}(x)$ in the direction perpendicular to the boundary 
and take into account that they must satisfy the boundary condition 
$\phi_{k_x}\left(\frac{a}{2}\right)=\lambda\partial_x\phi_{k_x}\left(\frac{a}{2}\right)$ (\ref{bound_is}) in order to ensure vanishing of the particle current. For an infinitely high 
barrier ($\lambda\to 0$), this boundary condition becomes a "hard wall" one and hence 
$\phi_{k_x}(x)=(2/L)^{1/2}\sin k_x(x-a/2)$ with $L$ being the length of the system. For a high enough
barrier, we can still use these eigenstates in the second line in equation (\ref{rho1}) since the derivative of $\sin k_x(x-a/2)$ at the boundary is finite. 
$E_{M,m}(k_x,{\bf k})=\frac{\hbar^2(k^2_x+k^2)}{2m}\mp\frac{\Delta}{2}$ is 
the electron spectrum in the Stoner model with $\Delta$ meaning the exchange-induced spin-splitting.    
Calculating the integral in Eq. (\ref{rho1}), one finds
\begin{eqnarray}
&&
\rho_{rM,m}(E_F,{\bf k})=\frac{\lambda^2m}{\pi \hbar^2}
(k^2_{M,m}-k^2)^{1/2}\Theta(k_{M,m}-|k|),
\nonumber\\
&&
k^2_{M,m}=2m(E_F\pm\Delta/2),
\label{rho2}
\end{eqnarray}
where $k_M$ and $k_m$ are the Fermi momenta of the majority and minority electrons, respectively, and 
$\Theta(x)$ is a step-function. The calculation of the convolutions (\ref{MM})--(\ref{Mm}) of the 
local electron spectral densities is now straightforward:
\begin{eqnarray}
&&
\rho_{MM,mm}=\frac{2\pi m^2A}{\hbar^4}(\lambda k_{M,m})^4
\label{MM1}\\
&&
\rho_{Mm}=\frac{2\pi m^2A}{\hbar^4}(\lambda k_{m})^4\times
\label{Mm1}\\
&&
\times\left[
\frac{\gamma^{1/2}(\gamma +1)}{2}-
\left(\frac{\gamma -1}{2}\right)^2
{\rm arccosh}\frac{\gamma +1}{\gamma -1}
\right],
\nonumber\\
&&
\gamma =k^2_M/k^2_m >1.
\nonumber
\end{eqnarray}  
The same approach can be used to calculate the total spin $S_z$ (\ref{Sz}) of the itinerant electrons penetrating into the barrier. 
At zero temperature  $|S_z|$ can be expressed in terms of the barrier parameter $\lambda$, the Fermi momenta of the majority ($k_M$) and minority ($k_m$) electrons, 
and the junction area $A$ as follows 
\begin{eqnarray}
|S_z|=\frac{\lambda^3 A(k^5_M-k^5_m)}{30 (2\pi)^2}.
\label{Sz1}
\end{eqnarray}  
%


\end{document}